\documentclass[
twocolumn,a4paper,superscriptaddress,aps,showpacs]{revtex4}
\usepackage{graphicx} 
\begin{document}
\def\la{\langle}
\def\ra{\rangle}
\def\om{\omega}
\def\Om{\Omega}
\def\vep{\varepsilon}
\def\wh{\widehat}
\def\tr{\rm{Tr}}
\def\da{\dagger}
\newcommand{\beq}{\begin{equation}}
\newcommand{\eeq}{\end{equation}}
\newcommand{\beqa}{\begin{eqnarray}}
\newcommand{\eeqa}{\end{eqnarray}}
\newcommand{\intf}{\int_{-\infty}^\infty}
\newcommand{\into}{\int_0^\infty}
%

\title{Bounds and enhancements for the Hartman effect}
\author{J. G. Muga}
\affiliation{Departamento de Qu\'\i mica-F\'\i sica,
Universidad del Pa\'\i s Vasco, Apdo. 644, Bilbao, Spain} 
\author{I. L. Egusquiza}
\affiliation{Fisika Teorikoaren Saila,
Euskal Herriko Unibertsitatea,
644 P.K., 48080 Bilbao, Spain}
\author{J. A. Damborenea}
\affiliation{Fisika Teorikoaren Saila,
Euskal Herriko Unibertsitatea,
644 P.K., 48080 Bilbao, Spain}
\affiliation{Departamento de Qu\'\i mica-F\'\i sica,
Universidad del Pa\'\i s Vasco, Apdo. 644, Bilbao, Spain}
\author{F. Delgado}
\affiliation{Departamento de Qu\'\i mica-F\'\i sica,
Universidad del Pa\'\i s Vasco, Apdo. 644, Bilbao, Spain}
\affiliation{Fisika Teorikoaren Saila,
Euskal Herriko Unibertsitatea,
644 P.K., 48080 Bilbao, Spain}

\date{June 26, 2002}
\begin{abstract}
The time of passage of the transmitted wave packet in a tunneling
collision of a quantum particle with a square potential barrier
becomes independent of the barrier width in a range of barrier
thickness. This is the Hartman effect, which has been frequently
associated with ``superluminality''.  A fundamental limitation on the
effect is set by non-relativistic ``causality conditions''. We
demonstrate first that the causality conditions impose more
restrictive bounds on the negative time delays (time advancements)
when no bound states are present. These restrictive bounds are in
agreement with a naive, and generally false, causality argument based
on the positivity of the ``extrapolated phase time'', one of the
quantities proposed to characterize the duration of the barrier's
traversal.  Nevertheless, square wells may in fact lead to much larger
advancements than square barriers.  We point out that close to
thresholds of new bound states the time advancement increases
considerably, while, at the same time, the transmission probability is
large, which facilitates the possible observation of the enhanced time
advancement.
\end{abstract}
\pacs{PACS: 03.65.-w}

\maketitle              

\section{Introduction}
The Hartman effect occurs when the time of passage $t_d$ of the
transmitted wave packet in a tunneling collision of a quantum particle
with an opaque square barrier becomes essentially independent of the
barrier width $d$ \cite{Hartman62,Fletcher85} ($t_d$ could be defined
as the time of passage of the peak, or by means of some average of
arrival or detection times as in Eq. (\ref{tbout}) below).  Since the
velocity, when defined by comparing the instants of the incoming and
outgoing peaks, may exceed arbitrarily large numbers, and there is a
related effect for photons or electromagnetic waves, this ``fast
tunneling'' has been frequently interpreted as, or related to, a
``superluminal effect'', see
e.g. \cite{EN93,MRRAR95,JOR98,Chiao98,Nimtz98}, even though, of
course, the velocity of light $c$ plays no role in the non-relativistic Schr\"odinger equation \footnote{The definition of the
entrance instant is more problematic than that of the outgoing one,
because of the interference between incident and reflected components
of the wave function.  Alternative entrance instants may be defined by
using the time of passage of the free motion wave packet across the
position of the left barrier edge, by averaging time with the positive
flux \cite{OR92}, through the read-out time of the half-width, or with
some other methods, which lead to similar results.}.

B\"uttiker, Landauer, and other authors have stressed however that
there is no physical law that turns peaks into peaks, namely, there is
no necessary ``causal'' link between the peak (or any other wave
packet feature) at the left barrier edge and the one at the right edge
\cite{BL82,LM94,Azbel94}.  In fact, even a well chosen statistical
ensemble of classical particles could show a similar ``superluminal
effect'' for certain barrier shapes. The ensembles can be actually
prepared, both in the classical and quantum cases, so that the
transmitted peak appears to the right even before the incident peak is
formed to the left \cite{DBM95}.

A related and controversial discussion is that of defining a
``tunneling time''.  Some of the definitions proposed lead in
tunneling conditions to very short times, which can even become
negative in some cases. This may seem to contradict simple concepts of
causality. The classical causality principle states that the particle
cannot exit a region before entering it. Thus the traversal time must
be positive. When trying to extend this principle to the quantum case,
one encounters the difficulty that the traversal time concept does not
have a straightforward and unique translation in quantum theory. In
fact for some of the definitions proposed, in particular for the so
called ``extrapolated phase time'' \cite{HS89}, the naive extension of
the classical causality principle does not apply for an arbitrary
potential, even though it does work in the absence of bound states, as
we shall justify below. Indeed, when bound states are present, e.g.,
for square wells, the advancements may be much more important (even
though also bounded by a causal principle) at low energies than those
characteristic of the Hartman effect for square barriers. Thus, it
seems appropriate to rename the time advancement due to the bound
states as an ``ultra Hartman effect''. We will also point out that
this enhanced effect may be readily observable since the low energy
time advancement is accompanied by high transmission probabilities
close to the onset of a new bound state.

The Hartman effect and related superluminal effects have triggered
quite a number of works, both theoretical and experimental, workshops,
and even the attention of the mass media.  Many of these works have
discussed the relativistic or ``Einstein causality'' principle, i.e.,
the limiting role of the velocity of light in the transmission of
signals \cite{DL93,SKC94,HB94,Diener96,KW99,RFG00,Hegerfeldt01}, which
must be applicable to relativistic wave equations; the influence of
the different wavepacket regions (rear, front) in the transmitted
signal, also in the non-relativistic case \cite{SBM95}; the
attainability of a sensible signal to noise ratio in superluminal
experiments with a small number of photons
\cite{ARS98,SMBC00,KDWMC01}; or the role of the frequency band
limitation of the signals \cite{HN94,Nimtz99,BT98,RM95,MB00}. Much
less attention has been paid to the consequences of the more primitive
and general causality principle stating that ``the effect cannot
precede the cause''.  In the context of scattering by spherical
potentials this basic causality principle implies analytical
properties of the $S$ matrix elements which impose certain limitations
on the possible time advancement of the outgoing wave packet
\cite{Nussenzveig72}.

The aim of this work is to describe bounds and enhancements for the
Hartman effect derived from the causality principle.  This principle
has been barely discussed in the context of one dimensional
collisions, with the exception of \cite{KF87,DK92,Sassoli94}. We
follow the track of these results by translating to the 1D case some
previous classical work by van Kampen \cite{vanKampen53II},
Nussenzveig and others for three dimensional collisions of spherical
potentials \cite{Nussenzveig72}.

Another type of limitation on the Hartman effect is that, for large
enough barriers, the above-the-barrier components with momentum
$p>p_b=(2mV_0)^{1/2}$, $V_0$ being the barrier energy, start to
dominate, so that the time of passage becomes ``classical'' and
depends again on the barrier width $d$.  This has been discussed by
various authors \cite{BSM94,HB94}, first of all by Hartman himself
\cite{Hartman62}, but we shall present a brief review for completeness
in the following section, which also introduces the Hartman effect
itself from a quantitative point of view \footnote{For limitations of
superluminal propagation due to quantum fluctuations in systems with
inverted atomic population see \cite{ARS98,SMBC00,KDWMC01}; the
Hartman effect is also affected by dissipation or absorption
\cite{RS94,NSB94}}. Section III presents the known causality bounds and
introduces the concepts and notation to establish the bound for the
absence of bound states in Section IV. Section V provides some
examples and shows the enhancement of the advancement that can be
achieved with square wells.  The article ends with a discussion and
some technical appendices.
\section{The Hartman effect and its large-barrier-width limitation}
We shall first review, briefly, the main features of the Hartman
effect and its large-$d$ limitation.  For a minimum-uncertainty wave
packet, the critical width $d_c$ separating the Hartman and classical
regimes may be estimated by the formula \cite{BSM94,DM96}
\beq
\label{crit}
d_c\approx \frac{\hbar}{4\Delta_p^2}
\left(\frac{(p_b-p_0)^3}{p_b+p_0}\right)^{1/2},
\eeq
where $p_0$ is the central momentum of the incident wave packet, and
$\Delta_p$ its standard deviation in the momentum representation.  This
limiting width can be obtained by equating the contributions to the
total transmittance above and below the barrier \cite{DM96}.

A simple derivation of the Hartman effect is based on the stationary
phase approximation. Let us write the transmitted wave function as
\beq
\psi_T(x,t) = \frac1{\sqrt{h}}\into dp\, e^{ixp/\hbar-iE_pt/\hbar+i\Phi_T}
\phi_{\rm in}(p,0) |T(p)|,     
\eeq
where $\phi_{\rm{in}}$ is the incident wave function (assumed to have
only components of positive momenta), and $T$ the complex transmission
probability amplitude, $T=|T|\exp(i\Phi_T)$.

If the initial state is narrowly peaked around $p_0$, the integral
will be appreciably different from zero only if the phase of the
exponential function is stationary near $p=p_0$. This implies a
``spatial delay'' with respect to the free-motion wave packet,
\beq
\Delta x= \hbar\frac{d\Phi_T}{dp}\bigg|_{p=p_0},
\eeq
and a corresponding ``time delay''
\beq
\label{deltat}
\Delta t(p_0)=\frac{\hbar m}{p_0}\frac{d\Phi_T}{dp}\bigg|_{p=p_0}.
\eeq
Taking into account the explicit expression of the transmission
amplitude $T$ for the square barrier, the total ``extrapolated phase
time'' (defined as the free motion term for crossing the barrier
width, $md/p_0$, plus the time delay) is easily shown to tend to a
constant for large $d$,
\beq\label{tfase}
\tau^{Ph}=md/p_0+\Delta t(p_0)\sim 2m\hbar/(p_b p_0),\;\;d\to\infty. 
\eeq
Of course this simple argument alone does not provide the whole
picture: it fails to predict the transition to the classical regime
described by Eq. (\ref{crit}).  Technically, the stationary phase
approximation leading to Eq. (\ref{deltat}) is inadequate for large
enough barriers because of the dominance of a different critical
point, namely, the ``barrier momentum'' $p_b$.

An alternative, more detailed approach is based on evaluating the
average passage instant in terms of the flux.  Throughout the paper we
shall assume that the barrier is located between $-a$ and $a$, the
barrier width being thus $d=2a$.  The average passage time at $a$ may
be defined as follows \cite{BSM94},
\begin{widetext}
\beq
\label{tbout}
\langle t\rangle_a^{out} = \frac{1}{P_T}\intf J_T(a,t) t\,dt=\quad{{m}\over{P_T}}\int_0^{\infty}\frac{dp}{p}
\,|\phi_{\rm in}(p,0)|^2\,|T(p)|^2\left[a
-x_0+\hbar\Phi'_T(p)\right],
\eeq
where 
\beq
x_0=x_0(p)\equiv-\hbar\,{\rm Im}\left(\frac{d\phi_{\rm in}(p,0)/dp}
{\phi_{\rm in}(p,0)}\right),
\eeq 
(for a Gaussian wave function $x_0$ becomes a constant, the center of
the packet at $t=0$), $J_T$ is the flux calculated with $\psi_T$, and
$P_T$ is the total (final) transmission probability for the wave
packet.
\end{widetext}
This result does not require the assumption of a narrow packet in the
momentum representation. A physical interpretation of $\la
t\ra_a^{out}$ as an average detection time is not straightforward,
since the flux $J_T$ is not a positive definite quantity, even for
wave packets composed entirely by positive momenta \cite{BM94}.  One
can show however that these ``averages'' do coincide with the ones
calculated with the positive definite ``ideal'' time-of-arrival
distribution of Kijowski \cite{Kijowski74,ML00}. They are also in
essential agreement with averages computed with localized detectors
modeled by complex potentials \cite{MBM95,MPL99}.

Formally, the integral represents an average, weighted by $|\phi_{\rm
in}(p,0)|^2|T(p)|^2/P_T$, of the time
$t^{Ph}(x_0,a)=m[a-x_0+\hbar\Phi_T']/p$ associated with each
momentum. If we subtract from this quantity the time required for a
classical particle (or a freely moving wave packet) to travel from
$x_0$ to the left barrier edge $-a$, one obtains again the
``extrapolated phase time'' of Eq. (\ref{tfase}).  The simple
asymptotic behavior of Eq. (\ref{tfase}) as $d\to\infty$ does not
however translate necessarily to $\la t\ra_a^{out}$, since for
sufficiently large $d$ the transmission factor $|T|^2$ is so strongly
suppressed at $p_0$ that the integral cannot be dominated by $p_0$;
instead, it becomes dominated by the region around $p_b$. Moreover,
the explicit integration shows other interesting deviations from the
simple ``constant-with-$d$'' behaviour of the monochromatic limit:
$\la t\ra_a^{out}$ in fact decreases slowly as $d$ increases due to
the filtering effect of the barrier \cite{DBM95}.  The total decrease
may actually attain a point where the difference between $\la
t\ra_a^{out}$ and a hypothetical free motion entrance instant
$t^{in}=|x_0+a|/p_0$ becomes negative \cite{DBM95}.

The ``extrapolated phase times'' for traversal should not be
over-interpreted as actual traversal times \cite{LA89,HS89}. Not only
because the quantization of the classical traversal time does not lead
to a unique quantity \cite{BSM94}, but also because a wave packet
peaked around $p_0$ is very broad in coordinate representation, so it
is severely deformed before the reference instant $t^{in}$, and at
$x=-a$ there is an important interference effect between incident and
reflected components \footnote{We could try to avoid the
interpretational pitfalls of the extrapolated phase time and look
instead at the time $\la t\ra_a^{out}$ for a wave packet initially
localized near the left edge of the barrier, and with a small spatial
width compared to the barrier length $d=2a$. In this way one might
identify the entrance time and the preparation instant at $t=0$ with a
tolerable small uncertainty. However, Low and Mende speculated
\cite{LM91}, and then Delgado and Muga showed \cite{DM96}, that this
localization implies the dominance of over-the-barrier
components. Similar conclusions are drawn from a two detector model
(one before and one after the barrier) when the detector before the
barrier localizes the particle into a small spatial width compared to
$d$ \cite{PMBJ97}.}.
\section{Negative delays}
\begin{figure}
\includegraphics[height=6cm]{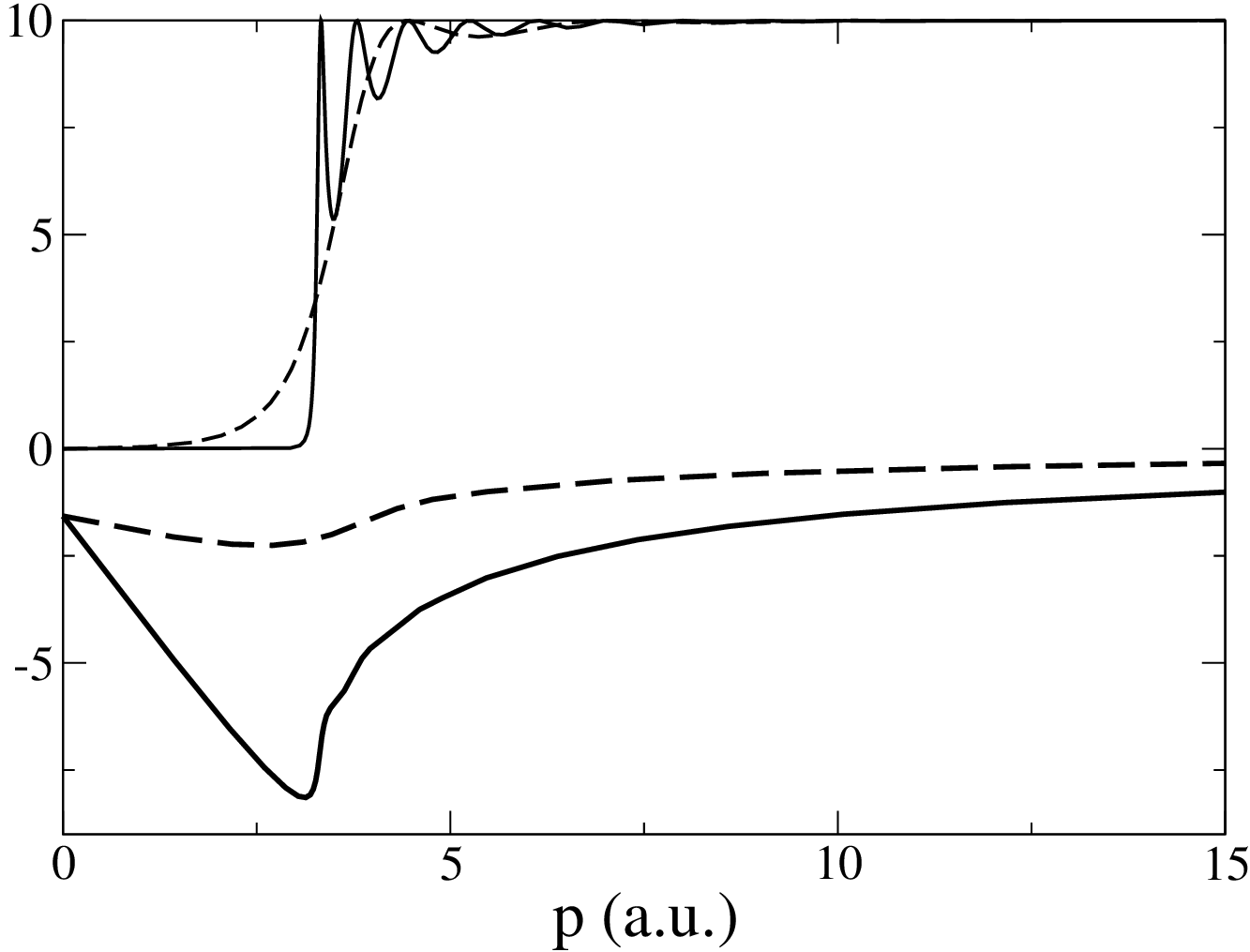}
\caption{\label{tdelayph}
$\Phi_T$ (lower curves) and $10\times |T|^2$ (upper curves) versus
momentum for a square barrier of ``height'' $V_0=5$ and for two
different widths, $d=1$ ({\it dashed lines}), and $3$ ({\it solid
lines}).  $m=1$. All quantities in atomic units}
\end{figure}

In partial wave analyses of three dimensional collisions with
spherical potentials, the time delay has been used mainly as a way to
characterize resonance scattering. One of the standard definitions of
a resonance is a jump by $\pi$ in the eigenphaseshifts $\delta_j$ of
the $\bf S$ matrix.  In one dimensional collisions the time delay has
also been used frequently to characterize non-resonant tunnelling,
where it may become negative. In fact the different delay signs
associated with the two types of effects, resonances and tunnelling,
are not independent.  In 3D it was soon understood by Wigner
\cite{Wigner55} that the increases and decreases of the phase should
balance each other. Since Levinson's theorem imposes a fixed phase
difference from $p=0$ to $\infty$, there must be intervals of negative
delay to compensate for the phase increases associated with the
resonances. A similar analysis applies in 1D to the transmission
amplitude.  In Figure \ref{tdelayph}, the phase of the transmission
amplitude for a square barrier is shown versus $p$ for two different
values of the barrier width $d$.
As $d$ increases, the scattering resonances ``above the barrier''
$p>p_0=(2mV_0)^{1/2}$ become more dense, and the $\pi-$jumps are
better defined, because of the approach of the resonance poles in the
fourth quadrant to the real axis. The corresponding increases of the
phase are compensated by a more negative delay in the tunneling
region.

Negative delays also arise if a pole of $T(p)$ crosses the real axis
upwards, when varying the interaction strength, to become a loosely
bound state in the positive imaginary axis.  Levinson's theorem
\cite{Sassoli94},
\begin{equation}\label{lete}
\Phi_T(0)=\cases{\pi(n_b-{1}/{2})
&if $\,T(p=0)=0$\cr
\pi n_b,&if $\,\,T(p=0)\neq 0$\cr}
\end{equation}
(the convention being that $\Phi_T(\infty)=0$, with $n_b$ denoting the
number of bound states), imposes then a sudden jump in the phase
$\Phi_T(0)$ that must be compensated by a strong negative slope.  This
effect is more important near threshold, i.e., when the pole is very
close to the real axis \cite{DK92}.  Similar effects have been
described for non-bound state poles in complex potential scattering
\cite{MP98}.

Note that the effects mentioned so far (due to resonances and bound
states) apply to arbitrary barriers.  Universal bounds on the allowed
negative delays may also be found.  Whereas positive delays can be
arbitrarily large, negative delays are restricted by ``causality
conditions'' \cite{Nussenzveig72}.  Some back-of-the-envelope
causality arguments may however be misleading. For example, if the
total time $\tau_T^{Ph}(-a,a)$ is to be positive, the delay cannot be
more negative than the reference free time,
\beq\label{bound}
\Delta t>-\frac{md}{p},
\eeq
cf. \cite{GP90}.  In fact this bound may be violated, in particular at
low energy in the proximity of a loosely bound state. This should not
surprise the reader after our warning against an over-interpretation
of the extrapolated time $\tau^{Ph}_T(-a,a)$. The flaw in the argument
is the assumption of positivity of $\tau_T^{Ph}(-a,a)$ because of an
inadequate translation from the classical trajectory case.
Nevertheless, rigorous bounds have been established by Wigner himself
and various authors in 3D collisions, see
\cite{Martin81,Nussenzveig72} for review.  In 1D collisions the
following bounds hold for even potentials with finite support between
$-a$ and $a$
\cite{DK92,Sassoli94} ($k=p/\hbar$):
\beqa
\Delta t&\ge&\frac{m}{\hbar k}   
\left\{-d-\frac{1}{2k}[\sin(2ka+2\delta_0)-
\sin(2ka+2\delta_1)]\right\}
\nonumber\\
&\ge&
\frac{m}{p}\left(-d-\frac{1}{k}\right).
\label{DiKi}
\eeqa
which follow from the bounds for the derivatives of the phase shifts,
\beqa\label{del0p}
{\delta_0}'&>&-a-\frac{1}{2k}\sin[2(ka+\delta_0)]>-a-\frac{1}{2k},
\\
\label{del1p}
{\delta_1}'&>&-a+\frac{1}{2k}\sin[2(ka+\delta_1)]>-a-\frac{1}{2k},
\eeqa
where the prime indicates derivative with respect to $k$.  Sometimes
Eqs. (\ref{DiKi},\ref{del0p},\ref{del1p}) -or their 3D analogs- are
pictorially or intuitively described as the result of the classical
causality condition (``the traversal time must be positive'')
corrected by a term of the order of a wavelength, that takes into
account the wave nature of matter, see
e.g. \cite{Wigner55,KF87,Nussenzveig72}.  In fact the proof,
summarized in the appendix, is based on the positivity of the norm
inside the barrier region, $\int_{-a}^a dx\,\psi_j(x)^2>0$
($\psi_j(x)$ are the even and odd real -improper- eigenstates of the
Hamiltonian).  Alternatively, it may be obtained from the positivity
of the dwell time \cite{Nussenzveig72,Martin81,Sassoli94}, which is in
this case the true ``causality principle'' behind the bound. The dwell time in the case at hand is defined \cite{Buttiker83} as 
\beq
\tau_D(-a,a)\equiv\frac{m}{\hbar k}\int_{-a}^a dx |\psi_{j}(x)|^2.
\eeq
Note that, unlike the extrapolated phase time, this quantity does not
distinguish between transmitted and reflected particles, and its
positivity is directly implied by its definition.
 
In Eqs. (\ref{DiKi}), (\ref{del0p}) and (\ref{del1p}), $\delta_j$,
$j=1,2$, are the phase shifts of the eigenvalues,
$S_j=\exp(2i\delta_j)$, of the $S$-matrix,
\beqa
{\bf S}(p)\equiv\left(
\begin{array}{cc}
T(p)  &  R(p)\\
R(p) &  T(p)
\end{array}
\right), 
\label{smat}
\eeqa
where $R$ is the reflection amplitude.  Since we are dealing with
parity invariant potentials,
\beqa
\nonumber
S_0&=&T+R,
\\
S_1&=&T-R.
\label{s0s1}
\eeqa
According to the bound in Eq. (\ref{DiKi}) the negative delay may be
arbitrarily large for small enough momenta and may diverge at $p=0$,
as it occurs when a bound state appears when making the potential more
attractive \cite{DK92}. In the absence of bound states, however, the
time advancement is actually bound by Eq. (\ref{bound}), instead of
Eq. (\ref{DiKi}). Thus, whereas the experiments looking for large
traversal velocities have been frequently based on evanescent
conditions in square barriers (tunneling) \cite{RMFPNS91,EN93a,SKC93},
square wells may lead to much larger advancement effects for barrier
depths near the thresholds.

In the next section we will show the validity of Eq. (\ref{bound}) in
the absence of bound states for a generic class of potentials and for
all real $k$.  Sassoli de Bianchi, using Levinson's theorem, had
previously pointed out the validity of Eq. (\ref{bound}) in the absence of
bound states, but only in the limit $k\to 0$ \cite{Sassoli94}.

Later we will insist on the further time advancements due to bound
states in the specific setting of square wells.

\section{Bound without bound states\label{bwbs}}
The bound for the time advancement without bound states follows from
the ``canonical product expansion'' of the $S_j$.  Note first that for
cut-off potentials $T$ and $R$, and thus both $S_0$ and $S_1$, are
meromorphic functions of $k$ in the entire complex plane \cite{DT79}.
In the absence of bound states, $S_j$ may only have simple poles in
the lower half plane, and the corresponding zeros in the upper half
plane,
\beq\label{sj}
S_j=\pm e^{-2ika}\prod_n \frac{(k-k_{j,n}^*)(k+k_{j,n})}{(k-k_{j,n})
(k+k_{j,n}^*)}.
\eeq
Even more, each pole $k_{j,n}$ in the fourth quadrant goes hand in
hand with a twin pole at $-k_{j,n}^*$.  This is the most general form
compatible with the following properties of the $S_j$:
\beqa
S_j(k)S_j^*(k^*)=1,
\label{s1}
\\
S_j^*(k)=S_j(-k^*),
\label{s2}
\\
|S_{a,j}|\le 1\;\; ({\rm Im} k\ge 0),
\label{s3}
\eeqa
where 
\beq
S_{a,j}=e^{2ika}S_j.
\eeq
The first two are the the ``unitarity'' and ``symmetry'' relations
that follow from the reality of the potential, whereas the last one
follows (in the absence of bound states) from van Kampen's causality
condition \cite{vanKampen53II,Nussenzveig72}, namely from the fact
that the total probability of finding the particle outside any sphere
of radius $r\ge a$ cannot be greater than one.  An equivalent
formulation is that the outgoing probability current, integrated from
$-\infty$ to $t$, cannot exceed the integrated ingoing current by more
than the absolute value of the integral of the interference
(ingoing-outgoing) term.  Van Kampen arrived at this causality
condition noticing that the causality condition for the scattering of
the Maxwell field, which implies a maximum velocity $c$, does not
apply to the Schr\"odinger equation; moreover, no wave packets could
be built which propagate with a sharp front for any finite interval of
time.  The mathematical arguments leading to Eq. (\ref{s3}) are
lengthy but admit a direct translation to the two partial waves of the
one dimensional case.  That Eq. (\ref{s3}) is fulfilled may be checked in
any case in the examples considered below.

Taking the logarithmic derivative on both sides of Eq. (\ref{sj}) one
obtains,
\beq\label{deriv}
d\delta_j/dk=-a-\sum_n\left(\frac{1}{|k-k_{j,n}|^2}+\frac{1}{|k+k_{j,n}|^2}\right)  {\rm Im}\, k_{j,n},
\eeq
assuming that $k$ is real.  Since all the poles lie in the lower
half-plane the second term is positive. In the absence of bound
states, $\delta_j'\ge -a$ and, since $\Phi_T=\delta_0+\delta_1$, the
transmission time delay does indeed satisfy the bound of
Eq. (\ref{bound}).

A bound state of energy $-(\hbar K_b)^2/(2m)$ for the $j$-partial wave
implies for $S_j$ an extra factor $(k+iK_b)/(iK_b-k)$, with $K_b>0$,
which spoils the lower limit $-a$ for $\delta_j'$. The contribution of
these bound-state terms will however be negligible for large $k$.  If
there is just one bound state corresponding to a pole at $k=iK_b$, one
obtains, similarly to the 3D case \cite{Nussenzveig72},
\beq
\Delta t\ge -\frac{m}{p}\left(d+\frac{1}{K_b}\right),  
\eeq
which may be interpreted in terms of an increased size of the
scatterer due to the broad range of the loosely bound state.
\section{Examples}
The above properties of the $S_j$ can be easily exemplified with the
aid of analytically solvable models. We have in particular checked the
validity of Eq. (\ref{bound}) for the square barrier,
\beq
V_{sb}(x)=V_0\,\chi(-a,a),
\eeq
for $V_0>0$, where $\chi(-a,a)$ is the characteristic function for the
barrier region.
\begin{figure}
\includegraphics[height=6cm]{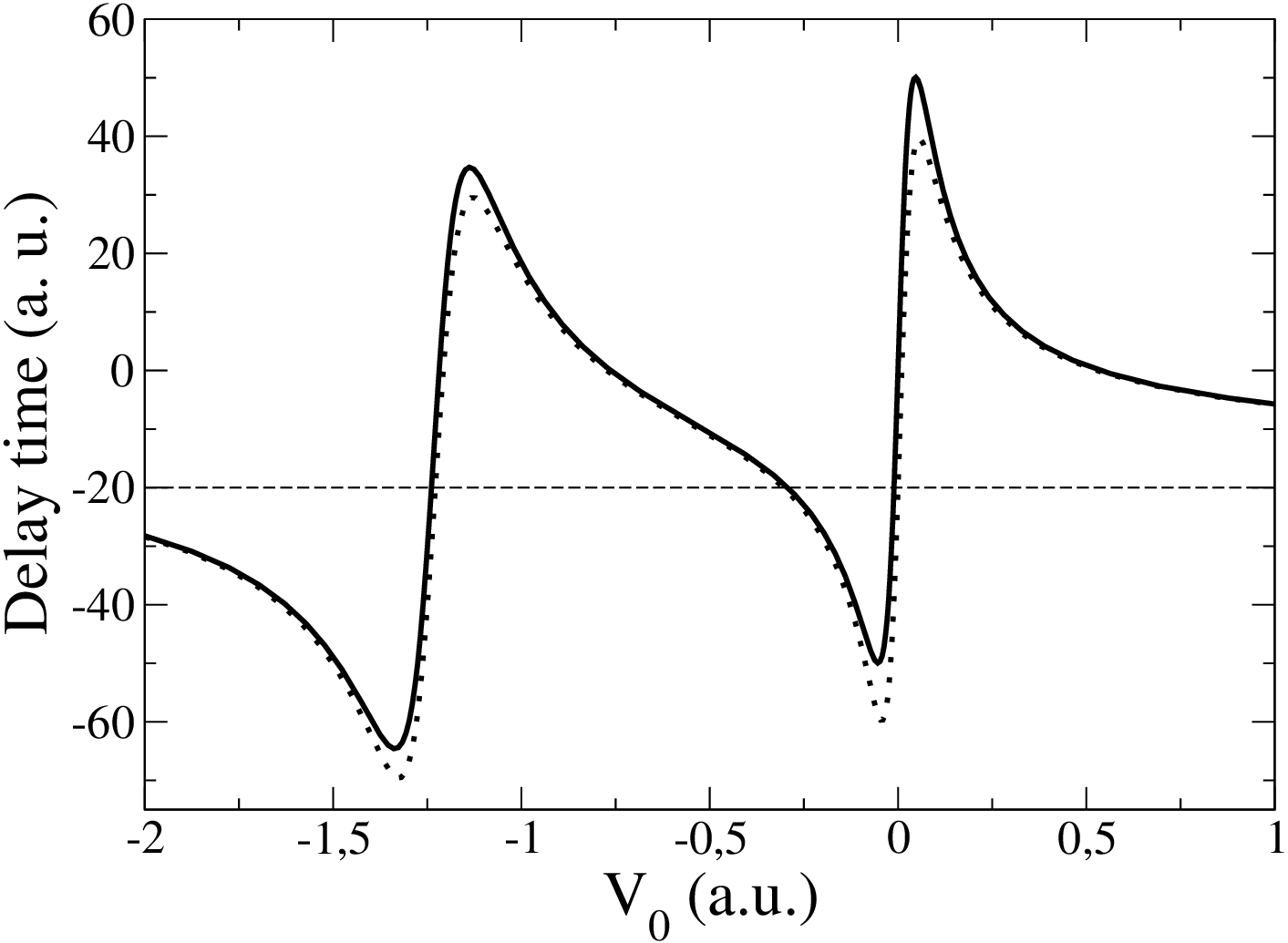}
\caption{\label{squaregraph} Time delay versus barrier height/depth
for a square potential. The incident momentum is $k=0.1$, mass $m=1$,
and barrier width $d=2$ (all magnitudes in atomic units).  The exact
expression for the time delay is depicted with a continuous line. The
dotted line corresponds to the top bound of (\ref{DiKi}), whereas the
dashed line is the constant quantity of the failed bound
(\ref{bound})}
\end{figure}
The eigenphaseshifts $\delta_j$ and their derivatives are easily
computed as explicit functions; however, those expressions are not
very enlightening for current purposes, so they are not displayed
here.  By making $V_0<0$ and allowing for the presence of bound states
the bound (\ref{bound}) is not satisfied any more, but
Eq. (\ref{DiKi}) does hold.

In order to make this explicit, we have plotted in
Fig. \ref{squaregraph} the delay time as a function of the well depth
for a set value of momentum, and, so as to compare, both constant
bound (\ref{bound}), and the more adequate Eq. (\ref{DiKi}). The line
depicting the constant value $-\frac{md}{p}$ is seen to cut the
computed line at points close to those that correspond to the onset of
a new bound state. On the other hand, the oscillatory bound of
Eq. (\ref{del0p}) keeps track of the injection of new bound
states. This effect takes place for each new bound state that comes
into play; in the figure only the first two thresholds for new bound
states appear, but the same structure repeats itself.  Note that, as
we are stressing throughout, the simple bound (\ref{bound}) does
indeed hold for barriers ($V_0>0$).

The analysis of the bound has pertained to stationary waves, or,
alternatively, to wavepackets highly centered in a particular momentum
component. It now behooves us to check whether the enhancement of time
delays can be carried over to wavefunctions more extended in momentum
space (and therefore more localized in space), with some chance of
being detected. For this purpose we will reexamine Eq. (\ref{tbout})
for the case of a square well. In such a situation $|T(k)|\sim O(k)$
as $k$ tends to zero, except when the depth of the well corresponds to
the threshold of a new bound state, that is to say, except when
$V_0=-(\hbar^2n^2\pi^2)/(8 m a^2)$. In other words, there is no
stationary state with zero energy in the presence of the well, but for
the exceptional threshold cases, when such a stationary state does
indeed exist, as it does for the free particle case. Therefore, except
for the exceptional threshold case, the integral presents no
singularity. On the other hand, for those exceptional situations
$|T(k)|\to1$ as $k$ tends to zero; since the time delay does not tend
to zero in that limit either, the integral will be divergent if the
asymptotic initial state in momentum representation does not go zero
fast enough. This should be no surprise: exactly the same divergence
would take place for free motion, and for exactly the same reason,
namely, the presence of $k=0$ stationary components, for which no
phase time can make sense.
\begin{figure}
\includegraphics[height=6cm]{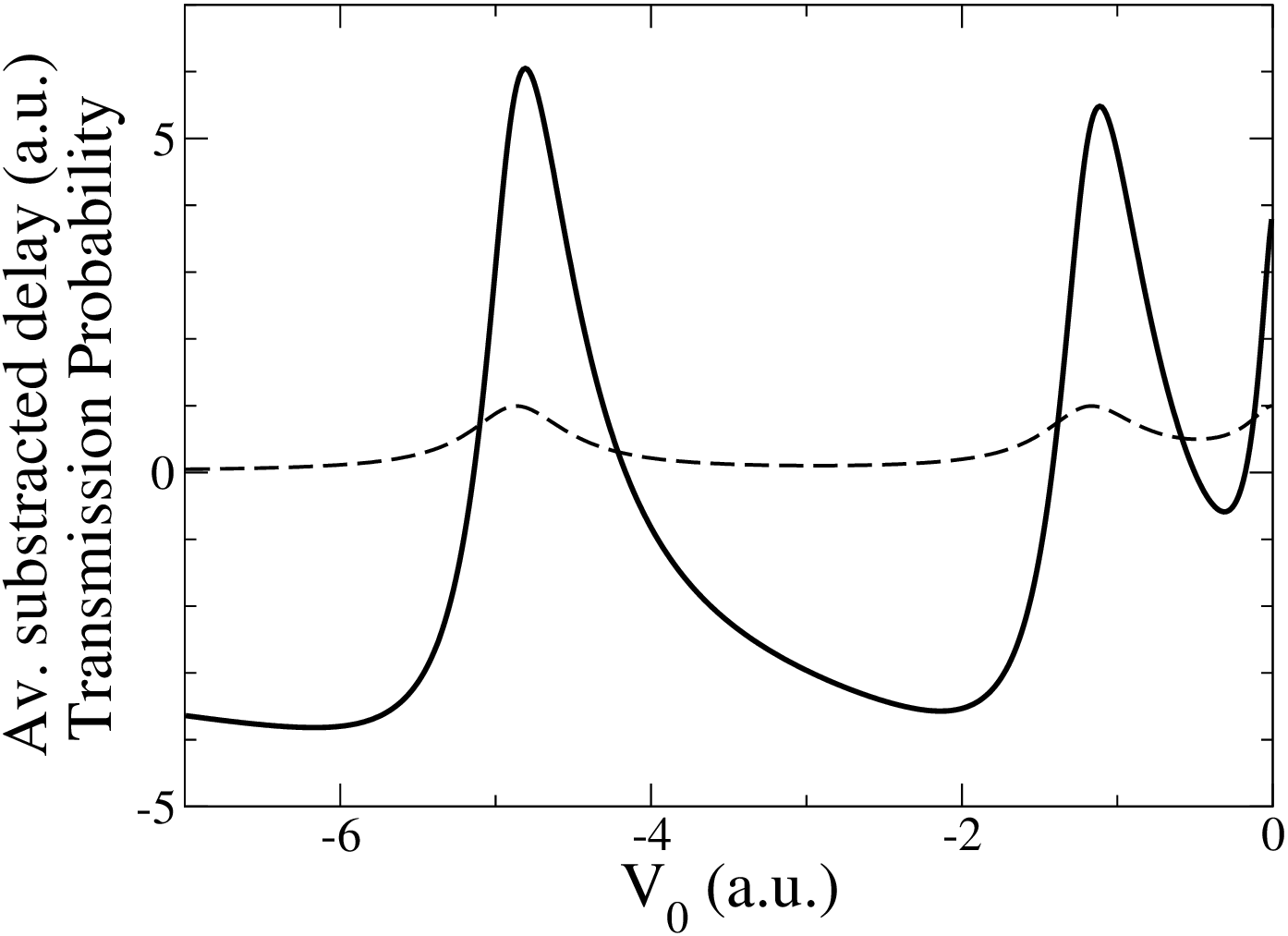}
\caption{\label{figinte} 
Transmission probability (dashed line) and averaged substracted
passage time (continuous line) versus potential for a truncated
gaussian wavefuncion with central momentum $k_0=\pi/8$ and dispersion
$1.0$, whose center is located at $x_0=-41$ at time $t=0$. The mass of
the particle is $m=1$ and the width of the square well/barrier is
$d=2$. All magnitudes in atomic units.}
\end{figure}
We depict in Figure \ref{figinte} the transition probability as a
function of the height/depth of a square barrier/well for an
asymptotic wavepacket which is gaussian in momenta, truncating out the
negative momentum part, and is centered at $x_0=-41a$ at the instant
$t=0$.  For the same wavepacket we also depict the ``averaged
substracted passage time'', that is, the result of integral
(\ref{tbout}), minus the time that a classical particle whose momentum
is the central momentum of the wavepacket would take to reach the
right-hand side of the potential well, from an initial position at
$x_0=-41a$ (the classical particle would be moving in the potential
well, for adequate comparison) \footnote{This quantity corresponds to that
defined as $\tau^{HFF}_T$ by Leavens and Aers, but for a constant,
i.e. the traversal time implied by Hauge et al. \cite{HFF87}. The fact that we are substracting simply a classical
particle time, instead of comparing with some other quantum
computation, is due to the divergence of the average passage time for
the free (quantum) particle case.}.  The
applicability of Eq. (\ref{tbout}) is ensured by the large initial
distance.  Negative substracted passage times
are apparent for zones of the potential strength for which a new state
has just been injected into the bound sector. These negative passage
times would be enhanced if the gaussian were closer to $k=0$ in
momentum space, be it because the center of the wavepacket in momentum
space moved towards the origin, or because the width of the packet
increased. An important aspect, apparent from the figure, is that the
transmission probability is big in some zones of negative average
passage times. This is due to the fact that the module of the
transmission amplitude, $|T(k)|$, has a maximum very close to $k=0$ if
the gap from the highest lying bound state to the continuum spectrum
is small. Therefore, we can have at the same time high transmission
probability, and strongly enhanced negative time delays; the idea of
using parameters of the potentials very close to those introducing a
new bound state in order to measure negative delays immediately
springs to mind.
\section{Discussion}
We have shown that the time advancement of the transmitted wave packet
of a particle colliding with a potential barrier without bound states
is bounded, due to the causality condition of van Kampen, by the
simple prediction based on assuming the positivity of the extrapolated
phase time. This positivity does not hold in the presence of bound
states, in which case a different bound allows for large advancements
al low incident energy near depth thresholds where a bound state
appears (``ultra Hartman effect'').  We have also argued, and provided
explicit examples, that these large advancements are indeed
observable, since the transmission probability is large at low
energies near the thresholds.

The present approach is formally limited by the assumption of a
cut-off in the potential and one may wonder about its relevance for
arbitrary potentials, but it is clear that the difference between the
actual (non-cut-off) potential and a potential truncated at an
arbitrarily large distance from the center cannot be physically
significant.  Thus, even though the canonical form assumed for the
$S_j$ in the complex plane may fail for the non-cut-off potential, all
observable features associated with real and positive values of the
wave number will be essentially unchanged, in particular the delay
times and their bounds.
\appendix
\section{Proof of  bounds (\ref{del0p}) and (\ref{del1p})}
Eq. (\ref{DiKi}) may be proven by using the even and odd
eigenfunctions $\psi_{j}(x)$ of the Hamiltonian, for which the
boundary conditions are
\beqa
\lim_{x\to-\infty}\psi_0(x)
&=&\left(\frac{2}{h}\right)^{1/2}
\cos(-px/\hbar+\delta_0),
\nonumber\\
\lim_{x\to\infty}\psi_0(x)&=&\left(\frac{2}{h}\right)^{1/2}
\cos(px/\hbar+\delta_0),
\nonumber\\
\lim_{x\to-\infty}\psi_1(x)&=&\left(\frac{2}{h}\right)^{1/2}
\sin(px/\hbar-\delta_1),
\nonumber\\
\lim_{x\to\infty}\psi_1(x)&=&\left(\frac{2}{h}\right)^{1/2}
\sin(px/\hbar+\delta_1)
\label{outer}.
\eeqa
In particular we shall use the fact that $\int_{-a}^a dx\,\psi_j^2>0$.
We start by calculating the logarithmic derivative of $\psi_0(x)$ at
$x=a$ from the known expression for the outer region, see
(\ref{outer}),
\beq
L_a\equiv\frac{\psi_0'(x)}{\psi_0(x)}\bigg|_{x=a}=
-\frac{p}{\hbar}\tan(pa/\hbar+\delta_0)=-k \tan(k a +\delta_0),
\eeq
where $\psi'(x)=d\psi_0(x)/dx$, and henceforward the primes denote
derivative wih respect to $x$.  Taking the derivative of $L_a$ with
respect to $k$,
\beq
\frac{d\delta_0}{dk}=-\frac{1}{k}\frac{dL_a}{dk}
\cos^2(ka+\delta_0)-a-
\frac{1}{2k}\sin[2(ka+\delta_0)].
\eeq
The first term on the right hand side may also be written as
\beqa
-\frac{\cos^2(ka+\delta_0)}{k}\frac{dL_a}{dk} &=&
-\frac{h}{2k}\psi_0^2(a)\left[\frac{\psi_0'(x)}{\psi_0(x)}\right]_k\bigg|_{x=a}
=\cr &=&-\frac{h}{2k}\psi_0(a)\tensor\partial_k\psi_0'(a)=\cr
&=&-\frac{h^3}{8\pi^2m}\psi_0(a)\tensor\partial_E\psi_0'(a)\,.\nonumber\\
\eeqa
Here and in what follows, the subscript $k$ and $E$ are shorthand
notation for the derivatives with respect to $k$ and $E$,
respectively, whereas the derivative with respect to $x$ is denoted by
primes, and the symbol $g(x)\tensor\partial_yf(x)$ indicates
$g(x)\partial_yf(x)-(\partial_y g(x))f(x)$ .  Repeating the same
operations for $x=-a$ one finds that
\beqa
&&[\psi_{0,E}(-a)\psi_0'(-a)-\psi_0(-a)
\psi_{0,E}'(-a)]
\nonumber\\
&=&-[\psi_{0,E}(a)
\psi_0'(a)-\psi_0(a)\psi_{0,E}'(a)].
\label{b-b}
\eeqa
We shall now prove that this is a positive quantity.
Taking the derivative of the stationary Schr\"odinger equation
with respect to 
energy one obtains for real eigenfunctions of $H$
the identity \cite{Smith60}  
\beq
\psi(x)^2=-\frac{\hbar^2}{2m}\frac{\partial}{\partial x}
\left(\psi(x) \psi_{,E}'(x)
-\psi_{,E}(x) \psi'(x)\right),
\eeq
so that, using (\ref{b-b}),  
\beq
\int_{-a}^{a}dx\,\psi_0(x)^2=\frac{\hbar^2}{m}
\left(\psi_{0,E}(a)\psi_0'(a)-\psi_0(a)
\psi_{0,E}'(a)\right),
\eeq 
whence the positivity of $-\frac{1}{k}\frac{dL_a}{dk}\cos^2(ka+\delta_0)$ follows, and thus bound (\ref{del0p}) is seen to hold.

Carrying out similar manipulations for the odd wave function 
$\psi_1(x)$, and using $\Phi_T=\delta_0+\delta_1$, 
(\ref{DiKi}) is found as a consequence of the  
positivity of the probability to find the particle in
the barrier region. 

\begin{acknowledgments}
This work is supported by Ministerio de Ciencia y Tecnolog\'{\i}a,
The University of the Basque Country, and the Basque Government.
J.A.D. acknowledges financial support by the Basque
Government. F.D. acknowledges financial support by Ministerio de
Educaci\'on y Cultura.
\end{acknowledgments}

\end{document}